\documentstyle[prd,aps,psfig]{revtex}
\begin{document}
\draft
\newcommand{\limfunc }{\mathrm }

\twocolumn[\hsize\textwidth\columnwidth\hsize\csname
@twocolumnfalse\endcsname

\author{Sandro S. e Costa\\
{\small {\it Instituto de F\'\i sica Te\'orica - UNESP}}\\
{\small {\it R. Pamplona, 145 - CEP 01405-000 - S\~ao Paulo - SP - Brazil}}\\
{\small {\it e-mail: sancosta@ift.unesp.br}}}
\title{Some integrals ocurring in a topology change problem }
\date{February 7, 2000}
\maketitle

\begin{abstract}
In a paper presented a few years ago, De Lorenci {\it et al.} showed, in
the context of canonical quantum cosmology, a model which allowed space
topology changes. The purpose of this present work is to go a step further
in that model, by performing some calculations only estimated there for several compact manifolds of constant negative curvature, such
as the Weeks and Thurston spaces and the icosahedral hyperbolic space (Best space).

$\,\,$

\centerline{PACS numbers: 98.80.Hw, 02.40.-k}
\end{abstract}

\vskip2pc]

\section{Introduction}

A few years ago De Lorenci {\it et al.} \cite{DeLorenci} presented a model
of quantum cosmology which allowed space topology changes, having as main
idea the use of the ``conditional probability interpretation'' to establish
selection rules for the possible changes of topology; the wavefunctions
involved in the process were of the type 
\begin{equation}
\label{um}\Psi =\Psi \left( \alpha ,\beta ,\xi ,\phi \right) =A_k\left(
a,\phi \right) e^{\xi F_k}\,\,\,,
\end{equation}
where $\alpha $ and $\beta $ are appropriated canonical variables built upon
the more common set of spherical coordinates $\left( \chi ,\theta ,\varphi
\right) $, the scale factor $a$ and the curvature $k$; $\xi $ and $\phi $
are, respectively, a dust field describing a ``distribution of irrotational
dust particles'' and a scalar field, both representing the matter content of
the model; and $F_k$ is basically a numerical coeficient obtained by
integration of certain functions constructed upon the `value' $\chi _0\left(
\theta ,\varphi ;V^3\right) $ of the radial coordinate of the fundamental
polyhedron's boundary of the 3-dimensional manifold $V^3$, of curvature $k$,
considered, written explicitly as 
\begin{equation}
\label{dois}F_k=\frac a{2\pi \hbar m}\int_{V^3}\frac{\sin 2\sqrt{k}\chi
_0\left( \theta ,\varphi ;V^3\right) }{2\sqrt{k}}d\theta \sin \theta
d\varphi \,\,\,.
\end{equation}
The topology changes would occur at some value $\xi $ of the the dust field,
when $a=\overline{a}$ and $\phi =\overline{\phi }$, such that the
conditional probability of having $k=-1$, $0$ or $+1$ would be 
\begin{eqnarray}
\label{tres}P_c\left( k|\overline{a},\overline{\phi }\right) &=& \frac{\left|
\Psi \left( k,\overline{a},\overline{\phi }\right) \right| ^2}{\sum_{k=0,\pm
1}\left| \Psi \left( k,\overline{a},\overline{\phi }\right) \right| ^2}\\ \nonumber
&=&\frac{A_k^2\left( \overline{a},\overline{\phi }\right) e^{2\xi F_k}}
{ \sum_{k'=0,\pm 1} A_{k'}^2\left( \overline{a},\overline{\phi }\right) e^{2\xi F_{k'}}}\,\,\, . 
\end{eqnarray}
So, when $\xi \rightarrow \pm \infty $ one has one of the $P_c\left( k|%
\overline{a},\overline{\phi }\right) $ equal to one and the other two null,
depending upon the value of $F_k$.

In \cite{DeLorenci} the values of the functions $F_k$ were only estimated
for two different compact manifolds, the Poincar\'e dodecahedral space $D^3$%
, of positive curvature, and the hyperbolic icosaedral space $I^3$ (also
known as Best space), of negative curvature; since there the authors claimed
that ``it is not possible to calculate the $F_i$'s exactly'' for these
manifolds, the importance of the present work is in the exact calculation of
the functions $F_k$ for several compact manifolds of constant negative
curvature, including the cited $I^3$.

\section{Some calculus in compact manifolds}

The functions $F_k$, such as presented in equation (\ref{dois}), are
probably uncomputable since the specific form of the functions $\chi
_0\left( \theta ,\varphi ;V^3\right) $ are difficult, if not impossible, to
determine; however, one can simply establish the following limits for the $%
F_k$'s: 
\begin{equation}
\label{estimativa}4\pi \frac{\sin 2\sqrt{k}\chi _{\min }}{2\sqrt{k}}\leq 
\frac{F_k}{\left( a/2\pi \hbar m\right) }\leq 4\pi \frac{\sin 2\sqrt{k}\chi
_{\max }}{2\sqrt{k}}\,\,\,,
\end{equation}
where $\chi _{\min }$ and $\chi _{\max }$ are, respectively, the radii of
the inscribed and circumscribed circunference of the the fundamental cell of
the manifold in consideration. In \cite{DeLorenci} the functions $\chi _0$
appear after performing an ``integration with respect to the variable $\chi $%
'', using as interval of integration $\left[ 0,\chi _0\left( \theta ,\varphi
;V^3\right) \right] $; so, in order to obtain a numerical value for the
functions $F_k$, it is easy to see that one can start with the integral 
\begin{equation}
\label{efeka}F_k=-\frac a{2\pi \hbar m}\int_{V^3}\left[ \sin ^2\sqrt{k}\chi
-\cos ^2\sqrt{k}\chi \right] d\chi d\theta \sin \theta d\varphi \,\,\,.
\end{equation}

Noticing now that 
\begin{equation}
\label{cinco}dV=\frac{\sin ^2\sqrt{k}\chi }kd\chi d\theta \sin \theta
d\varphi 
\end{equation}
is simply the element of volume for the spatial part of a
Friedmann-\-Robertson-\-Walker metric, written in spherical coordinates,
there are two possible ways to follow, one plainer and the other a little
more sophisticated; in both, however, one needs to redefine the coordinates
and limits of integration used. So, the next step consists in the use of
cylindrical coordinates $\left( \rho ,\varphi ,z\right) $, related to the
spherical coordinates $\left( \chi ,\theta ,\varphi \right) $ by means of
the relations 
\begin{equation}
\label{nove}\left\{ 
\begin{array}{c}
\cos 
\sqrt{k}\chi =\cos \sqrt{k}\rho \cos \sqrt{k}z \\ \sin \sqrt{k}\chi \sin
\theta =\sin \sqrt{k}\rho 
\end{array}
\right. 
\end{equation}
or 
\begin{eqnarray}
\label{dez} d\chi ^2&+&\frac{\sin ^2\sqrt{k}\chi }k\left[ d\theta ^2+\sin
^2\theta d\varphi ^2\right] =\\ \nonumber
&& d\rho ^2+\cos ^2\sqrt{k}\rho dz^2+\frac{\sin ^2\sqrt{k}\rho }kd\varphi ^2\,\,\,. 
\end{eqnarray}

Now, one has the interval of integration $\left[ 0,\rho _0\left( z,\varphi
;V^3\right) \right] $ for the coordinate $\rho $; the expression for $\rho
_0\left( z,\varphi ;V^3\right) $ is easily obtainable, since is only a
matter of using trigonometrical identities in the plane, {\it i.e.}, in the
triangles that compose the faces of each tetrahedron in which the
fundamental polyhedron can be divided\footnote{%
For more information on trigonometric identities in non-euclidean spaces one
can see references \cite{Thurston} to \cite{Coolidge};\cite{Peebles} is a classical book of cosmology with one section on spherical trigonometry.} using the following
procedure:

\begin{itemize}
\item  for each face draw a geodesic line perpendicular to it, connecting it
to the center $A$ of the polyhedron, and crossing it or its plane in a point 
$B$ (this line $AB$ gives the height $z$ of the tetraedron);

\item  for each edge draw a geodesic line perpendicular to it and connecting
it to the point $B$ of the face to which the edge belongs, crossing the edge
or its extension in a point $C$;

\item  complete the tetrahedron with one of the two vertices of the edge,
naming it as $D$.
\end{itemize}

\noindent These steps will create some `negative' tetrahedra, covering also
regions outside the polyhedron, and some `positive', covering only regions
of the polyhedron, each one of them having four right-angled triangles, one
of which (named here $BCD$) is the base of the tetrahedron; integration on
the compact manifold represented by the polyhedron is the difference between
the sums of the integrations on all of the positive tetrahedra and the
integrations on all of the negative tetrahedra. 
\begin{figure}
\psfig{file=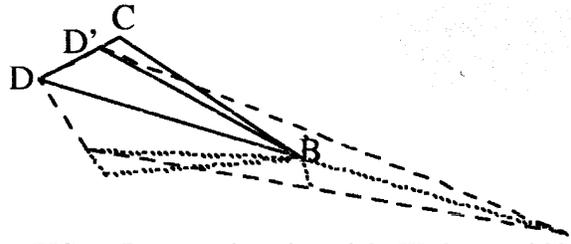,width=7.5cm}
\caption[a1]{Division of one face of the Weeks manifold 
(quadrilateral area bounded by the segmented lines) 
in several triangles;
the triangles labeled $BCD$ and $BCD'$ form, when connected to the center $A$ of 
the fundamental polyhedron, the basis of one 
`positive' and one `negative' tetrahedron, respectively. Notice that the point $A$, 
not shown in the figure, belongs to a different plane.}
\end{figure}

The easiest path of integration consists of simply making 
\begin{eqnarray}
\label{efeka2}F_k&=&-\frac a{2\pi \hbar m}\int_{V^3}\left[ 2\sin ^2\sqrt{k}%
\chi -1\right] d\chi d\theta \sin \theta d\varphi\\ \nonumber
&=&-\frac a{2\pi \hbar
m}\left[ 2kv-\int_{V^3}\frac{kdV}{\sin ^2\sqrt{k}\chi }\right]\,\,\,, 
\end{eqnarray}
where $v$ is the volume of the compact manifold where the integration is
being performed. The remaining integral in the right hand side of the last
equality must be done in the new set of cylindrical coordinates, where the
limits of integration for the particular case of negative curvature ($k=-1$)
are, in each tetrahedron\footnote{%
The trigonometric identities that lead to such result are showed in an
appendix at the end of this work.},
\begin{equation}
\label{ze}0\leq z\leq d_{AB}\,\,\,,\,\,\,0\leq \varphi \leq \stackrel{\wedge 
}{CBD}\,\,\,,
\end{equation}
and 
\begin{equation}
\label{doze}0\leq \rho \leq \rho _0\left( z,\varphi \right) =\mathrm {arctanh%
}\left[ \tan \stackrel{\wedge }{BAC}\frac{\sinh z}{\cos \varphi }\right]
\,\,\,.
\end{equation}
The integration in the coordinate $\rho $ is easily done and gives finally 
\begin{eqnarray}
\label{feum} &&\frac{F_{-1}}{a/2\pi \hbar m}=2v+\\ \nonumber
&&\int_0^{\stackrel{\wedge }{CBD}}d\varphi \int_0^{d_{AB}}\frac {dz}{\cosh ^2z}\ln \sqrt{\frac{\cos
^2\varphi +\tan ^2\stackrel{\wedge }{BAC}}{\cos ^2\varphi -\tan ^2\stackrel{\wedge }{BAC}\sinh ^2z}} 
\end{eqnarray}
from where numerical results can be obtained by plain numerical integration.
Notice that the same procedure can yield a formula for the volume of the
manifold.

Alternatively, one can start doing%
\begin{eqnarray}
F_k&=&-\frac{ak}{2\pi \hbar m}\int_{V^3}\left[ 1-\cot ^2\sqrt{k}\chi \right] 
dV\\ \nonumber
&=&-\frac{ak}{2\pi \hbar m}\left[ v+\int_{V^3}\nabla _\mu V^\mu dV\right]  
\end{eqnarray}
where $V^\mu $ is a vector satisfying the differential equation 
\begin{equation}
\label{seis}\nabla _\mu V^\mu =\left( \partial _\mu +\Gamma _{\mu \rho
}^\rho \right) V^\mu =-k\cot ^2\sqrt{k}\chi \,\,\,, 
\end{equation}
whose solution in spherical coordinates is\footnote{%
Here is used the identity
\par
$$
\frac{\pi ^2}{4m^2}\csc ^2\frac \pi m+\frac \pi {4m}\cot \frac \pi m-\frac 12%
=\sum_{k=1}^\infty \frac 1{\left( 1-k^2m^2\right) ^2} 
$$
found as equation $1.423$ of reference \cite{GR}.} 
\begin{eqnarray}
\label{sete}V^\chi &=&-\frac 12\left[ \sqrt{k}\cot \sqrt{k}\chi +k\chi \csc ^2%
\sqrt{k}\chi \right] \\ \nonumber
&=&-\frac 1\chi -\sum_{\ell =1}^\infty \frac{k^2\chi ^3}{\left( k\chi ^2-\pi ^2\ell ^2\right) ^2}\,\,\,, 
\end{eqnarray}
where the last equality was put to show clearly the behavior of the solution
when $k=0$. This result permits the use of Stokes's theorem \cite{Frankel}
to make 
\begin{equation}
\label{oito}-\int_{V^3}k\cot ^2\sqrt{k}\chi dV=\int_{S=\partial V^3}g_{\mu
\nu }V^\mu n^\nu dA\,\,\,, 
\end{equation}
where $n^\nu $ is a vector normal to the boundary $S$ of the fundamental
cell of the compact manifold $V^3$, obbeying the constraint $n^\mu n_\mu =1$.

In the procedure presented here the faces of the fundamental polyhedron that
represents a compact manifold appear, by construction, as surfaces of
constant $z$, allowing to use as element of area 
\begin{equation}
\label{onze}dA=\frac{\sin \sqrt{k}\rho }{\sqrt{k}}d\rho d\varphi \,\,\,.
\end{equation}

Finally, to carry out the integration the vector $V^\chi $ must be written
in cylindrical coordinates; only the component $V^z=V^\chi \partial _\chi z$%
, normal to the base of the tetrahedron, is important. This procedure can be
used also to give the volume of each tetrahedron, what allows to write, in
the case of negative curvature, 
\begin{eqnarray}
\label{quinze}&&\frac{F_{-1}}{a/2\pi \hbar m}=\\ \nonumber
&&\int_0^{\stackrel{\wedge 
}{CBD}}\frac{d\varphi}{\coth z} \ln \sqrt{\frac
{\cos ^2\varphi +\tan ^2\stackrel{\wedge }{BAC}}
{\cos ^2\varphi -\tan ^2\stackrel{\wedge }{BAC}\sinh ^2z}
}
\,\,\,, 
\end{eqnarray}
where $z=d_{AB}$.

\section{Numerical results}

To obtain numerical results the data -- volumes and coordinates of all
vertices for several hyperbolic compact manifolds -- contained in the
literature were used (see, for instance, \cite{Fagundes1} e \cite{Fagundes2}%
) together with those of the software {\it SnapPea}\footnote{{\it SnapPea}
is an electronic catalog of thousands of hyperbolic compact manifolds, each
one of them identified by volume and a code such as $m036\left(-3,2\right)$.}
\cite{SnapPea}; part of the data used are presented in Table \ref{table1}.
The manifolds choosed present in some way a degree of symmetry which
simplified the calculus, but, in principle, the approach followed can be
used to any compact manifold. All results are presented in the Table \ref
{table2} where they are compared with estimates done as in \cite{DeLorenci};
the result obtained for the Weeks manifold was used in \cite{CF1}. 
\vbox{
\begin{table}
\caption{Data for each manifold studied.
\label{table1}
             }
\begin{tabular}{cccc}
\bf Manifold     &  \bf volume &  $\chi _{min}$ & $\chi _{max}$  \\ \hline
\bf Weeks         &  $0.942707$  &$0.519162$&$0.752470$\\ 
\bf Thurston      &  $0.981369$  &$0.535437$&$0.748538$\\ 
\bf $m036(-3,2)$  &  $2.029883$ &$0.675646$&$1.014814$\\ 
\bf $m016(-4,3)$  &  $2.343017$  &$0.691286$&$0.895576$\\ 
\bf $m036(-2,3)$  &  $2.568971$  &$0.726205$&$0.895576$\\ 
\bf Best          &  $4.686034$   &$0.868298$&$1.382571$\\ 
\bf $v3469(+3,1)$ &  $5.137941$   &$0.808931$&$1.45241$\\ 
\end{tabular}
\end{table}
} 
\vbox{
\begin{table}
\caption{Summary of the results.
\label{table2}
             }
\begin{tabular}{cccc}
\bf Manifold     & $F_{-1}/\left( a/2\pi\hbar m\right)$ & $2\pi\sinh 2\chi _{min}$ & $2\pi\sinh 2\chi _{max}$ \\ \hline
\bf Weeks         & $9.28474$ &$7.76109$&$13.4518$\\ 
\bf Thurston      & $9.48385$ &$8.09029$&$13.3355$\\ 
\bf $m036(-3,2)$  & $13.4897$ &$11.3208$&$23.4987$\\ 
\bf $m016(-4,3)$  & $14.5526$ &$11.7314$&$18.3142$\\ 
\bf $m036(-2,3)$  & $15.3167$ &$12.6901$&$20.6181$\\ 
\bf Best          & $21.4948$ &$17.2847$&$49.6976$\\ 
\bf $v3469(+3,1)$ & $22.5418$ &$15.2178$&$57.1996$\\ 
\end{tabular} 
\end{table}
}

\section{Conclusion}

There are several formulations of quantum cosmology and the intention of
this work is to put some new light over a particular one, showing that the
wavefunctions built by the procedure of \cite{DeLorenci} present a
dependence on the volume of the compact manifold in consideration; aside
that, such wavefunctions have an additional dependence on the {\it shape} of
the fundamental cell of the manifold, due to a surface term that does not
appear in several other models.

To finish, it is also interesting to notice that the results presented here,
though of specific relevance for a particular model of quantum cosmology,
can be seen in a more generalized context, since this work presents a method
that allows to easily calculate the volume of the fundamental polyhedron of
a compact manifold. Explicitly, for the particular case of negative
curvature, the volume of each tetrahedron in which the fundamental
polyhedron can be divided is 
\begin{eqnarray}
\label{dezenove}&&v=\int_0^{\stackrel{\wedge }{CBD}}\frac{d\varphi }2
\times\\ \nonumber
&&\left\{ 
\frac{\limfunc{arctanh}\left[ \tanh z\sec \varphi \sqrt{\cos ^2\varphi +\tan
^2\stackrel{\wedge }{BAC}}\right] }{\sec \varphi \sqrt{\cos ^2\varphi +\tan
^2\stackrel{\wedge }{BAC}}}-z\right\} 
\end{eqnarray}
where again $z=d_{AB}$; alternatively,

\begin{eqnarray}
&&v=\int_0^{d_{AB}}\frac{dz}2\times \\ \nonumber
&&\left\{\frac{%
\mathop{\rm arctanh}
\left[ \tan \stackrel{\wedge }{CBD}\left( \cot ^2\stackrel{\wedge }{BAC}%
\mathop{\rm csch}
^2z-1\right) ^{-1/2}\right] }{\sqrt{\cot ^2\stackrel{\wedge }{BAC}%
\mathop{\rm csch}
^2z-1}}\right\}\,\,. 
\end{eqnarray}
These results must be compared with the more traditional ones given in \cite
{Coxeter1},\cite{Coxeter2} and \cite{Coolidge}.

\section*{Acknowledgments}

The author wants to thank the anonymous people of the Brazilian state of
S\~ao Paulo, who gave him financial support through grant n. 96/0052-3 of
the Funda\c c\~ao de Amparo \`a Pesquisa do Estado de S\~ao Paulo (FAPESP);
the author is also deeply grateful for the help of R.G. Teixeira and Prof.
Helio V. Fagundes.

\appendix

\section{Trigonometric identities}

In the non-euclidean geometry the trigonometric identities valid for a
triangle $XYZ$, with right angle $Z$, of sides $x$, $y$ and hypothenuse $z$
are \cite{Coxeter1} 
\begin{equation}
\label{dezoito}\sin Y=\frac{\sin \sqrt{k}y}{\sin \sqrt{k}z}\,;\,\,\cos Y=%
\frac{\tan \sqrt{k}x}{\tan \sqrt{k}z}\,;\,\,\tan Y=\frac{\tan \sqrt{k}y}{%
\sin \sqrt{k}x}\,\,. 
\end{equation}
Using the second identity in the right-angled triangle $BCD$, of right angle 
$C$, and the third one in the right-angled triangle $ABC$, of right angle $B$%
, one can write, for the tetrahedron $ABCD$ built as in the section 2, 
\begin{equation}
\label{vintum}\cos \stackrel{\wedge }{CBD}=\cos \varphi =\frac{\tan \sqrt{k}%
d_{BC}}{\tan \sqrt{k}\rho }=\frac{\tan \stackrel{\wedge }{BAC}\sin \sqrt{k}%
d_{AB}}{\tan \sqrt{k}\rho } 
\end{equation}
from where one obtains equation (\ref{doze}), after identification of $%
d_{AB} $ with $z$.

\end{document}